\documentclass[twocolumn]{aastex701}

\begin{document}

\title{Direct Imaging Constraints on Binary Planets and Exomoons around Epsilon Indi A b}

\author[orcid=0009-0004-7835-3356,sname='Garza']{Matson C. Garza}
\affiliation{Department of Physics, Massachusetts Institute of Technology, Cambridge, MA 02139, USA}
\email[show]{matgarza@mit.edu}  

\author[orcid=0000-0002-9521-9798,sname=Limbach]{Mary Anne Limbach}
\affiliation{Department of Astronomy, University of Michigan, Ann Arbor, MI 48109, USA}
\email{mlimbach@umich.edu}

\author[orcid=0000-0001-5831-9530,sname=Bowens-Rubin]{Rachel Bowens-Rubin}
\affiliation{Department of Astronomy, University of Michigan, Ann Arbor, MI 48109, USA}
\affiliation{Eureka Scientific Inc., 2542 Delmar Avenue, Suite 100, Oakland, CA 94602, USA}
\email{rbowens-rubin@mit.edu}

\author[orcid=0000-0003-1863-4960,sname=De Furio]{Matthew De Furio}
\affiliation{Department of Astronomy, The University of Texas at Austin, Austin, TX 78712, USA}
\email{defurio@utexas.edu}

\author[orcid=0000-0003-0593-1560,sname=Matthews]{Elisabeth C. Matthews}
\affiliation{Max Planck Institute for Astronomy, Heidelberg, Germany}
\email{matthews@mpia.de}

\author[orcid=0000-0003-4557-414X,sname=Franson]{Kyle Franson}
\affiliation{Department of Astronomy and Astrophysics, The University of California, Santa Cruz, Santa Cruz, CA 95064, USA}
\altaffiliation{NHFP Sagan Fellow}
\email{kfranson@ucsc.edu}

\author[orcid=0000-0003-3130-2282,sname=Millholland]{Sarah C. Millholland}
\affiliation{Department of Physics, Massachusetts Institute of Technology, Cambridge, MA 02139, USA}
\email{sarah.millholland@mit.edu}

\author[orcid=0000-0003-3904-7378,sname=Pearce]{Logan A. Pearce}
\affiliation{Department of Astronomy, University of Michigan, Ann Arbor, MI 48109, USA}
\email{lapearce@umich.edu}

\author[orcid=0000-0001-7246-5438,sname=Vanderburg]{Andrew Vanderburg}
\affiliation{Center for Astrophysics,  Harvard \& Smithsonian, 60 Garden Street, Cambridge, MA 02138, USA}
\email{avanderburg@cfa.harvard.edu}

\begin{abstract}

\noindent
Epsilon~Indi~A~b is a directly imaged $\sim$6\,$M_{\rm Jup}$ exoplanet orbiting a nearby (3.6\,pc) K-dwarf at $\sim$30\,AU. We analyze archival JWST/MIRI 15\,$\mu$m coronagraphic imaging of this planet to search for directly imaged satellites orbiting Eps Ind A b. Within the planet’s Hill sphere (radius $R_H \approx$ 2.3\,AU or 1.3\,$\lambda/D$), we compare single- and double-PSF models using Bayesian evidence. We find that a double-PSF (binary planet) fit is preferred. This apparent preference can most plausibly be explained by systematics, although follow-up observations would be required to fully rule out a binary planet interpretation. We construct a contrast curve of the exoplanet after removing this feature, demonstrating sensitivity to companions as faint as $0.03\times$ the F1550C flux of Eps~Ind~A~b (equivalent to T = 130 K, 1.3\,$M_{\rm Jup}$) at large separations ($>2$ AU). We also demonstrate sensitivity to brighter companions $0.2\times$ the F1550C flux of Eps~Ind~A~b (equivalent to T = 180 K, 2.5\,$M_{\rm Jup}$) down to separations of 0.52\,AU (1.3\,pixels; 0.29\,$\lambda/D$; 144\,mas). 
This study demonstrates that JWST/MIRI can directly detect exomoons or binary planets inside the Hill sphere of directly imaged exoplanets orbiting neighboring stars.  

\end{abstract}


\section{Introduction} 

Over the past decade, interest in the search for binary planet systems and exomoons has grown steadily, yielding a few intriguing candidates but no confirmed detections \citep{Teachey2018_Kepler1625,Heller2019,Kreidberg2019,Teachey2020,Kipping2022,Heller2024}.
 Multiple indirect techniques have been proposed for the identification of planetary companions. One direction focuses on the effects of such moons on the transit signals of short-period planets. These include transit timing variations \cite[TTVs;][]{Simon2007}, transit duration variations \cite[TDVs;][]{Kipping2009}, and spectral exo-Io signatures around hot Jupiters \citep{Oza2019}. Such efforts have demonstrated that the exomoon occurrence rate is low around short period planets down to the scale of Galilean analogs \citep{Teachey2018}. This is in agreement with theoretical models, which predict the occurrence rate to be much larger for wider-orbit planets \citep{2018MNRAS.480.4355C,2020MNRAS.499.1023I}. 
 
 When searching for exomoons orbiting these wide-orbit exoplanets/free floating planets, additional methods of detection are available. A joint microlensing survey with Roman and Euclid would be capable of detecting exomoons around free floating planets \citep{Bachelet2022}. 
 When an exoplanet has been directly imaged, exomoons can be found by adapting the traditional star-planet detection methods to the planet-moon systems such as
 Doppler spectroscopy \citep{Vanderburg2021, Ruffio2023, Horstman2024}, astrometry \citep{Winterhalder2025}, and the transit method \citep{Limbach2024}. 
 
In principle, giant exomoons can also be directly imaged like their host planets. This approach was applied by \citet{Lazzoni2020} to VLT/SPHERE observations of 27 brown dwarfs and planets, yielding one candidate secondary companion in the DH Tauri system at a separation of 10~AU. A significant difficulty with direct imaging as an exomoon detection method is that the small angular separations of exomoons often run into the diffraction limit. However, in some cases, this limit can be overcome. Infrared direct imaging has proved successful in detecting close brown dwarf binaries below the diffraction limit using empirically derived PSF models \citep{DeFurio2019, DeFurio2022, Calissendorff2023, DeFurio2025}, providing a potential path forward to directly imaging exomoons or binary planets at closer separations to directly imaged exoplanets. 

\object[eps Indi b]{Epsilon Indi A b} is the nearest confirmed exoplanet directly imaged to date \citep{Matthews2024, Matthews2026, Sanghi2026}. This nearly solar-aged (${\sim}3.5$ Gyr) super-Jupiter ($\sim$6--8~$M_\mathrm{J}$) is relatively well-isolated from its star at a wide periastron ($\sim$17~AU), providing ample dynamically stable space that can be probed for a binary planet or exomoon companion; at this system distance, 1 AU corresponds to a full 0.27{\arcsec} of projected separation. In combination with its brightness (magnitude 11.20 in F1550C), these factors make it our best opportunity to probe an exocircumplanetary space via direct imaging. This paper explores the potential of JWST/MIRI to accomplish this. Using the F1550C data of \citet{Matthews2024}, we first search for companions of Eps~Ind~A~b by adapting double-PSF fitting methods \citet{DeFurio2023, DeFurio2025} to MIRI coronagraphic imaging. We then perform injection recovery in order to build contrast curves. This allows us to put the first constraints on the parameter space of potential binary planets and exomoons in the Eps~Ind~A~b system.

\begin{deluxetable*}{cccc}

\tablecaption{Literature parameters for Eps Ind A b. \label{tab:epsilon indi ab parameters}}
\tablehead{
\colhead{Parameter} & \colhead{Value} & \colhead{Reference}
}
\startdata
Mass (dynamical) & $ \rm 6.31^{+0.60}_{-0.56} $ $\rm M_{Jup}$ & M \\
Mass (photometric) & $\rm {\sim}8.6$ $\rm M_{Jup}$ & M \\
Semi-major axis & $\rm 28.4^{+10}_{-7.2}$ AU & M \\
Eccentricity & $\rm 0.40^{+0.15}_{-0.18}$ & M \\
Inclination & $\rm 103.7 \pm 2.3^\circ$ & M \\
Temperature & $\rm{\sim}275$ K & M \\
Age & $\rm 3.5^{+0.8}_{-1.0}$ Gyr & C \\
App. mag. (F1065C) & 13.16 & M \\
App. mag. (F1550C) & 11.20 & M \\
Separation & $\rm 4.114 \pm 0.010$\arcsec & M \\
Position angle & $\rm 37.39 \pm 0.43^\circ$ & M \\
\enddata
\tablecomments{Separation and position angle are measured relative to the host star Eps Ind A based on data taken on 2023-07-03. References: M = \citet{Matthews2024}, C = \citet{Chen2022}.}
\end{deluxetable*}

Our understanding of Epsilon Indi as a system has grown significantly over the past decades. The host star, Epsilon Indi A, is a $0.782 \pm 0.023~M_\odot$ K-dwarf \citep{Lundkvist2024} at a nearby distance of $3.6384 \pm 0.0013$ pc. In the early 2000s, a distant T-dwarf binary was discovered \citep{Scholz2003, Volk2003, McCaughrean2004}. The first indications of Eps~Ind~A~b appeared in 2002 as a radial velocity trend \citep{Endl2002}. In subsequent years, additional radial velocity and astrometry observations bolstered the case for its presence \citep{Zechmeister2013, Kervella2019, Kervella2021, Feng2019, Feng2023, Philipot2023}, yet direct imaging of the planet remained elusive \citep{Geissler2007, Janson2009, Viswanath2021, Pathak2021}. This changed when \citet{Matthews2024} (GO 2243) successfully used JWST/MIRI coronagraphy to obtain a positive detection in 2023. The properties of Eps~Ind~A~b are summarized in Table \ref{tab:epsilon indi ab parameters}.

\section{Observations} \label{sec:observations}

For this study, we use JWST/MIRI coronagraphy imaging of Eps~Ind~A~b from JWST GO 2243, employing the data products reduced by \citet{Matthews2024}; the details of the reduction process are described therein. These observations were taken using the F1550C mid-infrared filter, which uses a four-quadrant phase mask \citep[4QPM;][]{Rouan2000, Rouan2007} to suppress starlight. This filter was chosen for our analysis because the lower-mass, cooler objects we are searching for would have very high contrast with their host planets in shorter-wavelength bands like F1065C. Eps~Ind~A received 1 dither for a total time of 3922 s, while the reference star DI~Tuc received 5 dithers for a total time of 13622 s. All observations were taken sequentially with the FASTR1 readout pattern on 3 July 2023 (following standard background observations to remove JWST ``glow stick'' stray-light artifacts). More observational details are given in Extended Data Table 1 of \citet{Matthews2024}; all GO 2243 data may be obtained from the MAST archive at \dataset[10.17909/70tb-g166]{https://doi.org/10.17909/70tb-g166}.

Importantly, these data have already had residual PSF starlight from Eps Ind A subtracted using Karhunen-Loeve Image Processing \citep[KLIP;][]{Soummer2012} with the nearby reference star DI Tucanae. For our purposes, we chose to use the data with 15 KL modes in PSF subtraction to optimally minimize the presence of residuals. Despite this processing, some artifacts originating from the 4QPM phase boundaries persist in a diagonal line rotated approximately 18 degrees counterclockwise from the vertical. These artifacts are near, but not coincident with, the exoplanet.

\begin{figure*}[h!]
\plotone{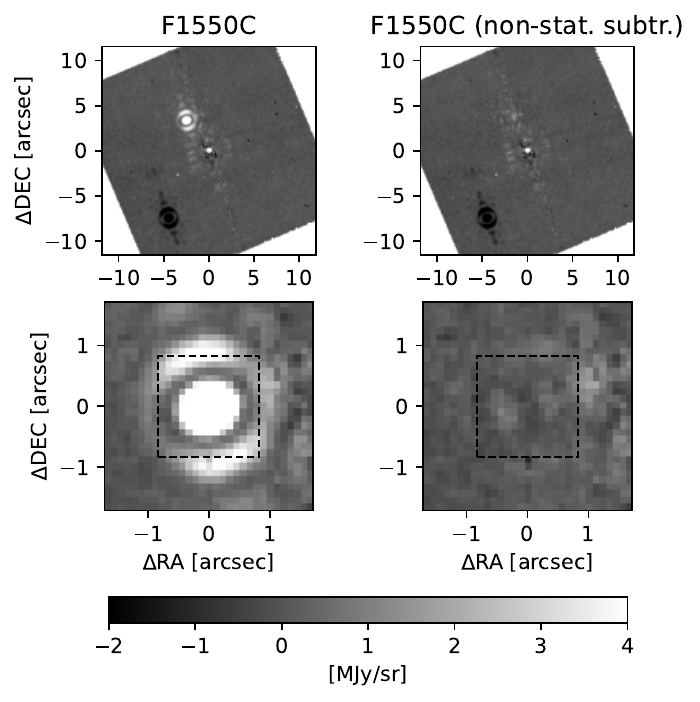}
\caption{North-aligned Stage 3 F1550C data of Eps Ind A before and after a non-statistical single-PSF subtraction. The second row panels are centered on the location of Eps~Ind~A~b and represent the 31$\times$31 pixel cropped regions; the dashed box indicates the 15$\times$15 pixel region used in fitting. The prominent negative source in the lower left quadrant is Gaia DR3 6411654761473726464, a visual companion to DI Tucanae and our reference star for planetary PSF subtraction. 
\label{fig:stage3}}
\end{figure*}

\begin{figure*}[h!]
\plotone{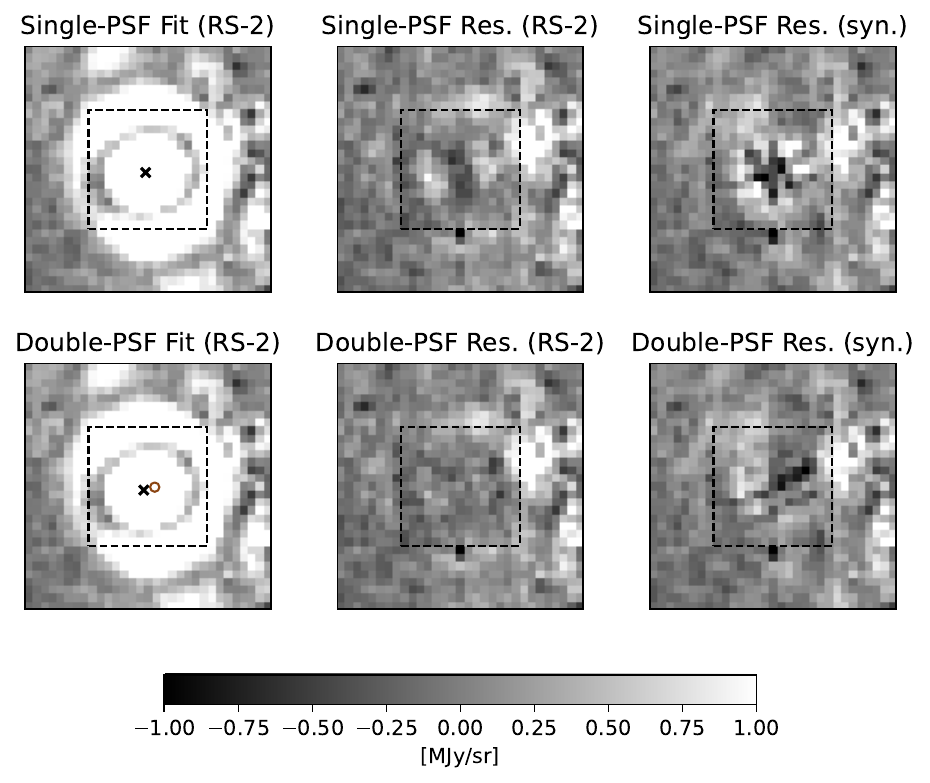}
\caption{Fits and residuals of stage 3 F1550C data subtracted using the results of \texttt{PyMultiNest} single- and double-PSF fitting. Each panel indicates the 31$\times$31 pixel cropped regions, while the dashed box indicates the 15$\times$15 pixel test region used in fitting. The residuals are significantly reduced when a double-PSF fit is applied. In the left two panels, the injected data are shown, with a black cross and brown circle indicating the fitted locations of the primary and secondary, respectively (using RS-2 as a PSF model). The residuals of these fits are shown in the middle panels. The right panels show the residuals of fits using a synthetic \texttt{StPSF} model; these residuals are visibly larger than those of the reference star model.
\label{fig:comparison}}
\end{figure*}

\section{Methods} \label{sec:methods}

\subsection{Determination of search radius}

Our goal is to conduct a direct imaging search for exomoons/binary planets around Eps~Ind~A~b and to put constraints on their parameter space. We restrict our analysis to the dynamically stable region in which a companion could form and remain gravitationally bound to the planet. Such companions must orbit within a fraction of the planet’s Hill sphere \citep{Nesvorny2003}, which we adopt as the outer boundary of our search region.

We estimate the Hill radius of the planet, given by
\begin{equation}
    r_{H}=a(1-e)\left(\frac{m}{3(M+m)}\right)^{1/3}
\end{equation} where $a$, $e$, $m$, and $M$ are the planet semi-major axis, planet eccentricity, planet mass, and stellar mass, respectively. Using the planetary parameters from Table \ref{tab:epsilon indi ab parameters} and a stellar mass of 0.782 $M_\odot$ \citep{Lundkvist2024}, we find $r_H$ to be 2.3 AU, which at a distance of 3.64 pc corresponds to 640 mas (5.8 pixels). 

This value approaches the $\lambda / D$ diffraction limit of F1550C, which is 492 mas (4.49 pixels). However, recent work has demonstrated that companions are detectable within a fraction of the diffraction limit of MIRI using double-PSF fitting \citep{DeFurio2019,DeFurio2025} and Kernel Phase Interferometry \citep{Adelman2025} around sources of similar mid-IR brightness (brown dwarfs, free-floating planets and white dwarfs) to Eps~Ind~A~b. We therefore adopt the double-PSF fitting technique to search for binary planets around Eps~Ind~A~b.

\subsection{Fitting routine}
\subsubsection{Construction of a PSF model and likelihood statistic}

The first step of companion detection is to obtain or generate a reference PSF for reference differential imaging \cite[RDI;][]{SmithTerrile1984, Lafreniere2009} of the planet. Fortunately, a suitable secondary reference star was present in the observations of the reference star, DI Tucanae: this star was identified by \citet{Matthews2024} as Gaia DR3 6411654761473726464 (hereafter referred to as RS-2). For the purposes of this paper, we used only this single dither as our PSF model; during fitting, we performed subpixel alignment using the bicubic interpolation routine of \texttt{scipy.ndimage.shift}. Both the target image and reference image were cropped to 31$\times$31 pixels, centered on Eps~Ind~A~b and RS-2 respectively.

In order to put constraints on any undetected bodies orbiting Eps~Ind~A~b, we perform single- and double-PSF fitting using the \texttt{PyMultiNest} package \citep{Buchner2014} for nested sampling to compute the Bayesian evidence for both single- and double-PSF models. In order to run nested sampling, we need to specify a likelihood function for our test region. Correlated noise is present in both our target and reference images; importantly, non-Gaussian spatially correlated noise is induced near the 4QPM phase boundary. For simplicity, in our analysis we assume non-spatially correlated Gaussian noise with a uniform standard deviation $\sigma$ in our test region. This approximation is sufficient for our purposes; as described in Section \ref{subsec:success_criteria}, we do not rely on Bayesian evidence in our injection recovery success criteria. 

\texttt{PyMultiNest} requires a likelihood statistic in order to perform fitting. Under the above approximation, the likelihood statistic for data $D$ given model $M$ is
\begin{equation}
    \mathcal{L}(D|M)=\prod_i^N\frac{1}{\sqrt{2\pi\sigma^2}}\exp{\left[\frac{-r_i^2}{2\sigma^2}\right]}
\end{equation}
where $r_i$ represents the residual for a pixel $i$ in our test region and $N$ represents the total number of such pixels. It can be shown that this is directly related to the $\chi^2$ statistic via
\begin{equation}
    \ln\left(\mathcal{L}(D|M)\right) = \ln\left((2\pi\sigma^2)^{-N/2}\right) - \frac{1}{2}\chi^2
\end{equation}
(That is, maximizing this likelihood statistic is equivalent to minimizing the $\chi^2$ statistic.)

We must still provide a value for the standard deviation $\sigma$. To estimate this, we used non-statistical methods to subtract the reference image from the target image and directly take the standard deviation of the resulting residuals. We accomplished this by exploiting the following fact: a misaligned reference leads to an asymmetry due to oversubtraction on one side and undersubtraction on the other. Choosing our reference scaling so that the average pixel value in our test region is zero, this asymmetry can be quantified to leading order by introducing a photometric dipole moment $\vec{p}$:
\begin{equation}
    \vec{p}=\sum_i^Nr_i\vec{x_i}
\end{equation}
where $r_i$ represents the residual for a pixel $i$ in our test region and $\vec{x_i}$ represents the coordinates of this pixel. To align the reference image, we then used Newton's method to find the zeros of this function (i.e., to eliminate the leading-order residual asymmetry) with respect to x- and y-shifts of the reference image. This process produced the residuals shown in the right panel of Figure \ref{fig:stage3}.

\subsubsection{Model parameters}

Our three single-source PSF model parameters were: \begin{itemize} 
    \item \texttt{x} - The x-coordinate of the source relative to the center of the test region.
    \item \texttt{y} - The y-coordinate of the source relative to the center of the test region.
    \item \texttt{norm} - The brightness of the source relative to the reference PSF.
\end{itemize}
Our double-source PSF model includes three additional parameters to specify the properties of the secondary source:
\begin{itemize}
\item \texttt{sep} - The separation of the secondary from the primary.
\item \texttt{PA} - The position angle of the secondary relative to the primary.
\item \texttt{contrast} - The flux ratio between the secondary (satellite) and primary (planet; Eps Ind A b) sources.
\end{itemize}

\subsubsection{Prior distributions}

In both models, our priors for \texttt{x}, \texttt{y} were uniform from -5 to 5 pixels in F1550C; these were chosen to roughly encompass the Airy disk within which the primary must be found; for \texttt{norm}, our prior distribution was uniform from 0.5 to 1.0, as direct inspection of the data suggested RS-2 to be slightly brighter than Eps~Ind~A~b. (Indeed, our best fit value for \texttt{norm} turned out to be about 0.65).

In the binary model, our prior distribution for secondary separation \texttt{sep} was uniform in log-space from 0.1 to 6.5 pixels to encompass the Hill sphere. As no exomoons have been detected as of writing, this prior was chosen to reflect our lack of knowledge of exomoon system architectures and to treat different scales of separation on equal footing. A log-uniform prior also approximates the semimajor axis distribution for small, closely-orbiting planets \citep{Petigura2013}. We found that a linearly uniform separation prior gave consistent results. Our prior distribution for \texttt{PA} was uniform from 0 to $\rm2\pi$, while our prior distribution for \texttt{contrast} was flat in log-space from $10^{-2}$ to $10^0$ to treat all contrast scales on equal footing.

\subsection{Injection recovery}
\subsubsection{Success criteria}
\label{subsec:success_criteria}

Having completed our search for binary planets around Eps~Ind~A~b (results are in Section \ref{results}), we next determined the parameter space in which such companions could have been detected. We used an injection–recovery test that employs the same signal extraction and double-PSF fitting techniques as in our initial search.

For injection recovery, our target image is reused as our injection PSF. We consider the recovery of an individual injection (fixed contrast, separation, and position angle) as successful if the fitted location of the primary and the relative location of the secondary are within one pixel of their true values. Additionally, any fits that find a contrast greater than one are considered failures (this would imply that the fitting code misidentified the primary and secondary). With these criteria, we find that at the sensitivity limit, 68\% and 95\% of fits return a contrast within 0.24 and 0.62 magnitudes of the injected companion, respectively. This demonstrates that we successfully recover the flux of the injected companion to within $\sim$0.5~mag, in addition to its position.

We considered 20 equally-spaced separations from 0 to 6.50 pixels (not including 0). At each separation, we then considered eight equally-spaced angles ranging from 0 to 315 degrees with respect to the x-axis of the data. In order for a given contrast-separation combination to be considered detectable, it must be successfully recovered for a predetermined fraction of these angles. (E.g., for a 75\% success threshold, a given contrast-separation combination must be recovered at six out of the eight injection angles considered.) For each separation, there exists a maximum detectable contrast; to compute this, we first injected a companion at a contrast of 0.5, performing a binary search with 0 and 1.1 as the upper and lower bounds until the maximum detectable contrast was determined to a precision of $10^{-3}$. If the search returned a final maximum contrast value above 1 at a given separation, it was considered a failure (this case suggests that no companion would be detectable).

The false positive rate for a single injection at a specified contrast, separation, and angle can be estimated as the probability a random point within our prior space satisfies the primary success criterion (i.e., that the fitted location of the secondary is within one pixel of the injected location). Our false positive estimation is dependent only on the location of the secondary; conceptually, given that the code always returns some fit, we check how often a fitted companion location accidentally lines up with that of an injected companion. We simulated this numerically by drawing samples from our prior; for a single angle, we found that this probability ranges from 6\% to 0.15\% over a separation range of 1.6 to 6.5 pixels, respectively. As discussed in Section \ref{sec:results}, this is the range over which we were able to successfully perform injection recovery at all 8 angles. Requiring success at all angles means that the overall false positive probability for a given separation should be no greater than $0.06^8 \approx \mathcal{O}(10^{-8}) \%$, which is satisfactory for our purposes.

Importantly, this false positive rate estimation applies only to our injection recovery, where an injected location is known and we are checking whether the code consistently identifies it within one pixel. An actual candidate detection would likely have a much higher false-positive rate because vetting such a candidate requires statistical analysis, which was not included in this estimation. The code always returns a fit regardless of whether it is significant. Furthermore, the ability of bright artifacts to masquerade as significant signals would likely also increase the false-positive rate for candidate detections.

\subsubsection{Conversion to physical parameter space}

Once we have the minimum recoverable contrast at each separation, we compute the relative photometry of the minimum detectable companion by comparison with the measured flux of Eps~Ind~A~b. Since this quantity is the direct sum of the fluxes of our hypothetical primary and secondary, the secondary flux can be expressed as a function of the total flux and the contrast: 
\begin{equation}
    F_s = \frac{c}{1+c}F_{tot}
\end{equation} where $F_s$ is the hypothetical secondary's flux and $F_{tot}$ is Eps~Ind~A~b's measured flux. If a signal of contrast $c_0$ is subtracted off beforehand, this formula becomes
\begin{equation}
    F_s = \frac{c}{(1+c_0)(1+c)}F_{tot}
\end{equation}
We then convert the flux to temperature space using the assumption that the source is a 1 $\mathrm{R_J}$ blackbody, and then to mass space using the planetary evolution tables for temperature-mass conversion of \citet{Fortney2007}. Recently, \citet{Bowens-Rubin2025} found that cold (100-400\,K) giant planets' spectral energy distributions are well approximated by blackbodies at mid-IR wavelengths. Indeed, applying this flux-temperature-mass conversion method to Eps~Ind~A~b itself yields a reasonable mass estimate of $6.32~\mathrm{M_J}$, which consistent with the dynamical mass of  $6.31^{+0.60}_{-0.56}~\mathrm{M_J}$ \citep{Matthews2024}. 

\section{Results}\label{results}

\subsection{A Search for Binary Planets around Eps~Ind~A~b}
\label{subsec:search}

\begin{figure*}[h!]
\plotone{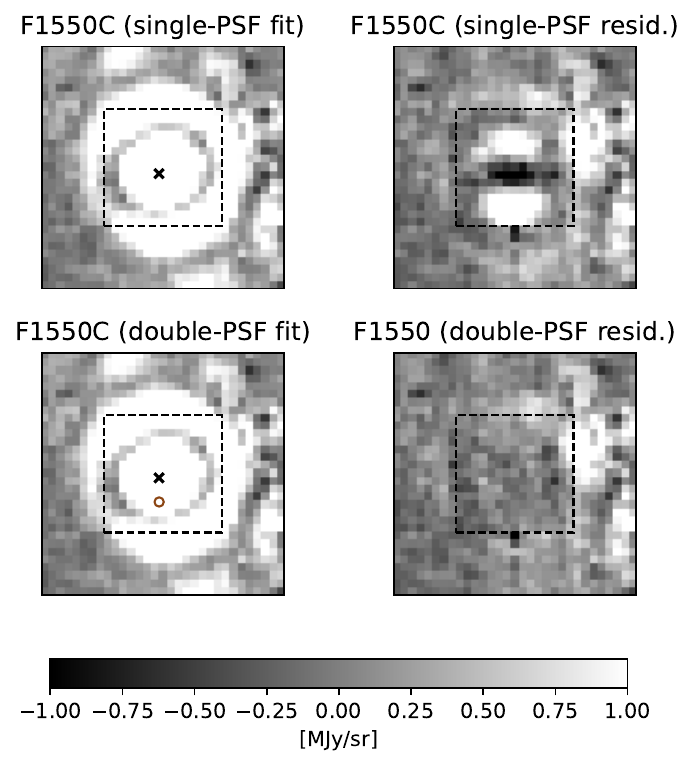}
\caption{Fits and residuals of stage 3 F1550C data with the detected signal removed and a fake companion injected 3 pixels below the primary at a contrast of 0.2 (similar brightness to the signal). The residuals were obtained via subtraction using the results of \texttt{PyMultiNest} single- and double-PSF fitting. The black cross and brown circle indicate the fitted locations of the primary and secondary, respectively. Each panel indicates the 31$\times$31 pixel cropped region, while the dashed box indicates the 15$\times$15 pixel test region used in fitting. Note the clear ``hamburger" residual pattern of under- and over-subtraction that occurs with a single-PSF fit (top-right panel); this pattern is aligned with the line connecting the primary and secondary. These residuals are drastically reduced when a double-PSF fit is applied. 
\label{fig:injection}}
\end{figure*}

Using this setup we conduct a search for companions to the directly imaged planet Eps~Ind~A~b using our single- and double-PSF fitting algorithm. The residuals from the results of our two fits are shown in Figure \ref{fig:comparison}; also shown in this figure are the residuals obtained by using a \texttt{StPSF} synthetic PSF model instead of a reference star. These residuals visually appear more chaotic than those of the reference star, which comparison of the single-PSF residual standard deviations confirms (approx. 0.40 and 0.31 MJy/str for the synthetic and reference star PSF models, respectively). As a result, we conclude that our use of RS-2, rather than a synthetic PSF, is preferred.

Having made this choice, we find that the double-PSF fit notably reduces the residuals in the image. The light-dark-light (undersubtraction, oversubtraction, undersubtraction) ``hamburger" pattern visible after the single-PSF subtraction is what one would expect if a companion were present, although this particular pattern is very faint. Figure \ref{fig:injection} shows an example of the same PSF fitting performed on a fake companion injected 3 pixels below the primary at a contrast of 0.2; note the alignment of the hamburger pattern with the position angle of the injected companion (180$^\circ$ for this injection).

\begin{figure*}[ht!]
\includegraphics[width=0.85\textwidth]{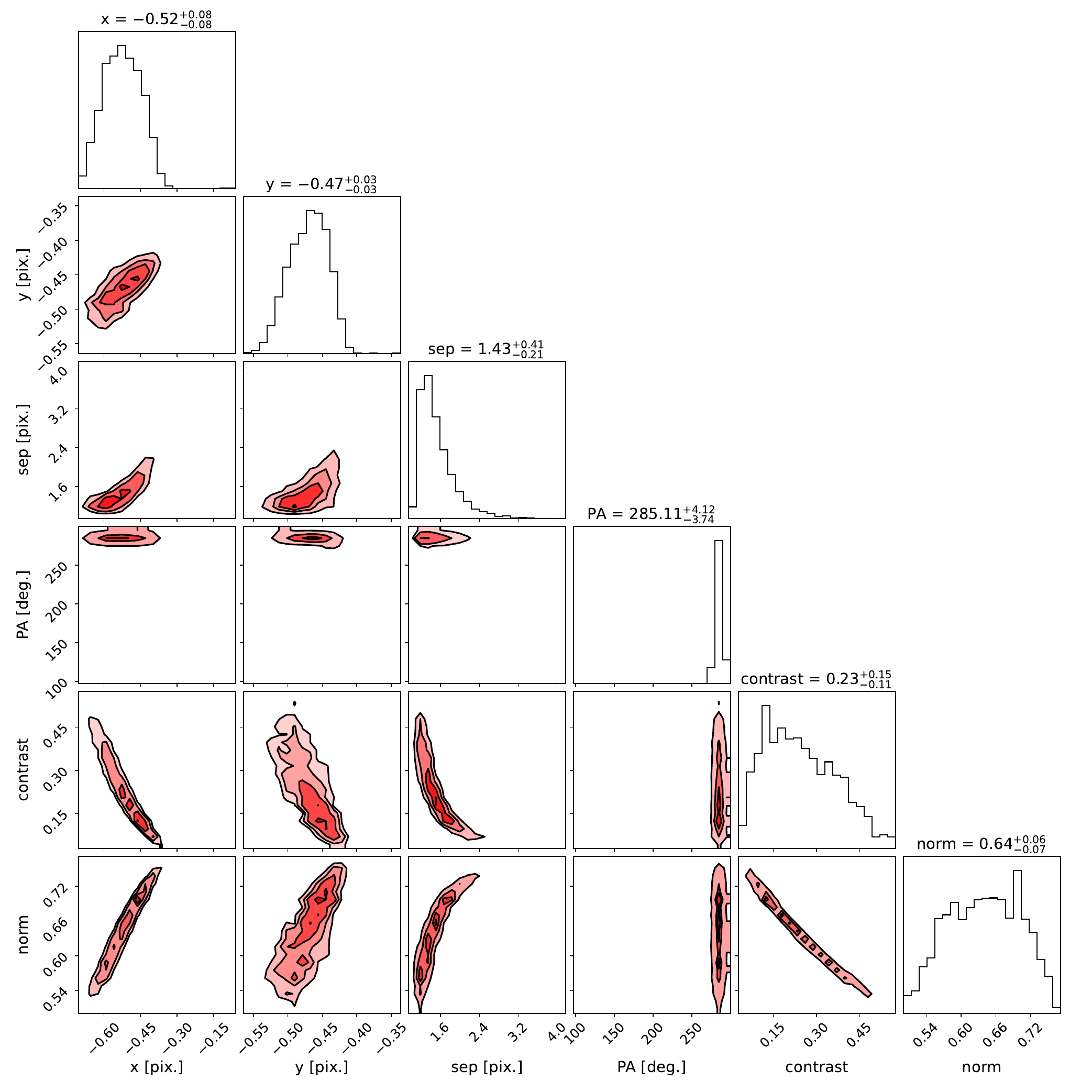}
\centering\caption{Corner plot for \texttt{PyMultiNest} double PSF fitting of the Eps Ind A b MIRI data showing the best-fit binary planet parameters relative to the known exoplanet.
\label{fig:corner}}
\end{figure*}

Following \citet{DeFurio2023}, we considered the binary model to be statistically favored when the natural logarithm of the evidence ratio $K=Z_{\rm binary}/Z_{\rm single}$ is greater than 5 \citep{Trotta2008}. In this case, the binary model was favored with $\ln(K) = 24.5$. The double-PSF fitting algorithm converged to a separation of $1.43 \pm 0.32$ pixels ($0.158 \pm 0.035$ \arcsec) and a position angle of $285.1 \pm 4.1$ degrees at a contrast of $0.23 \pm 0.13$. The results of the fit are shown in the corner plot in Figure \ref{fig:corner}.

However, despite the Bayesian evidence in favor of the double-PSF (binary planet) model fit to Eps~Ind~A~b, we have ample reason to be skeptical that this source represents a true companion rather than noise or a systematic artifact. 

First, the reduced $\chi^2$ values for the single- and double-PSF fits are 0.713 (222 d.o.f.) and 0.516 (219 d.o.f.), respectively. The fact that these are less than one is consistent with our crude estimate of the error per pixel, but could also indicate overfitting. Following \citet{Absil2011} and \citet{Gallenne2015}, we estimate the significance of the double-PSF's lower reduced $\chi^2$ to be 3.6$\sigma$. This alternative analysis does indicate marginal significance, but shows a much weaker preference than the Bayesian evidence would suggest. Additionally, our Bayesian evidence assumes random, uncorrelated noise, whereas the residuals are dominated by correlated speckle noise. As a result, our assumption of uniform, uncorrelated noise across our small test region is likely significantly overestimating the $\ln(K) = 24.5$ evidence ratio.

Second, we have only a single reference PSF (RS-2), which was observed on a different region of the detector. JWST/MIRI coronagraphic PSFs are known to vary across the field \citep{Perrin2012}, and the apparent signal may simply result from structural differences between the target and reference PSFs (e.g., brighter-fatter effect), producing a residual that mimics a companion.

We conclude that systematics provide a far simpler and more plausible explanation for the detected companion signal and are therefore the most likely cause. However, if this residual pattern were a binary planet, it would be a $T\sim190\,K$, 2.8 $M_{\mathrm{J}}$ source located at a projected separation of 0.58 AU from the primary (the Hill radius is around 2.5 AU). As shown in Figure \ref{fig:corner}, if this is the case, there is a significant degeneracy between possible combinations of separation and contrast since we are below the diffraction limit. We stress that due to the amount of potential confounding factors, we do not consider this a robust candidate companion at this time.

\subsection{Results from Injection Recovery Testing} \label{sec:results}
 
\begin{figure*}[ht!]
\includegraphics[width=0.85\textwidth]{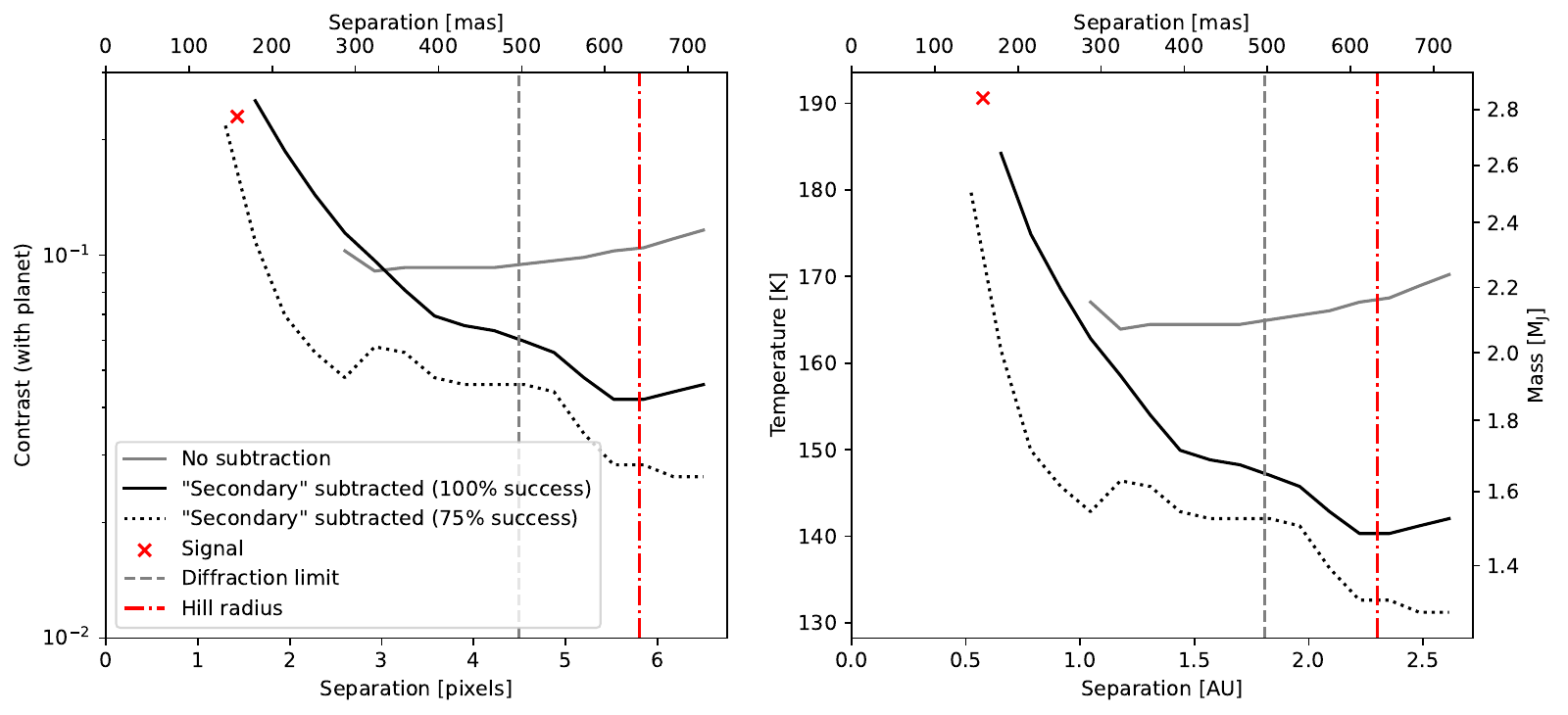}
\centering\caption{Contrast sensitivity limits to exomoons/binary planets around Eps~Ind~A~b for F1550C data. Once the signal described in Section \ref{subsec:search} is removed, our injection recovery testing demonstrates that we are capable of detecting binary planets as small as 1.3\,M$_{\rm J}$ (T = 130\,K) at the outer edge of the Hill sphere and 2.5\,M$_{\rm J}$ (T = 180\,K) at 0.52\,au. The companion signal detected in Section \ref{subsec:search} is marked by the red $\times$.
\label{fig:contrast curve}}
\end{figure*}

We successfully produced three contrast curves for the F1550C filter. One (``No subtraction") was built without removing the weak secondary signal discussed in Section \ref{subsec:search}, while the other two (``Secondary subtracted" at 100\% and 75\% success thresholds) were built after subtracting it from the original target image as if it were a point source. These are shown in Figure \ref{fig:contrast curve}. The former curve is cut off at a minimum separation of 2.5 pixels, as it approaches within 1 pixel of the detected signal and begins to interfere with recovery of injected signals. The latter two begin to run into problems as 1 pixel of separation is approached, reaching minimum separations of 1.625 and 1.3 pixels, respectively. This occurs because the fitting code often mistakes the companion for the primary at sufficiently low separations and contrasts. This situation is not helped by decreasing the contrast, and so these curves hit a sudden end using our methodology. The fact that our detected signal is so close to these cutoffs is further reason to treat its significance with healthy skepticism. Rigorous Bayesian evidence-based success conditions would likely be able to statistically determine binarity at even lower separations.

For reference purposes, the traditional diffraction limit (gray dashed line) and Hill radius (red dashed line) are also shown. The binary planet mass-separation space above the lines would have been detectable with this observation.

These contrast curves look qualitatively similar to those obtained for detections around stars: as the separation increases, the maximum contrast is set by the background noise limit. At close separations, PSF subtraction residuals limit the minimum detectable companion. For stars, the PSF subtraction effects often manifest as speckles which masquerade as planets. Here, however, PSF subtraction effects enter differently: the primary cause of the decrease in maximum contrast at low separations is a growing indistinguishability between the predictions of single-PSF and double-PSF models. The residuals of a binary system after the subtraction of a fitted single-PSF model become increasingly suppressed with decreasing separation.

We find that we successfully recover 100\% of injections if they are more than about 1.6 pixels apart, or 0.65 AU (at a maximum contrast of 0.25). At further separations, we find ourselves background-limited at a maximum contrast of around 0.03, which approximately corresponds to a 130 K object (1.3 $M_\mathrm{J}$). 

\section{Discussion} \label{sec:disc}

\subsection{The weak detection}

In Section \ref{subsec:search}, we conclude that systematics provide the simplest and most plausible explanation for the detected companion signal and are therefore the most likely cause. Additional mid-infrared data could confirm or reject this explanation; for instance, an astrophysical origin for the signal would produce an anomalous mid-infrared excess detectable in upcoming Cycle 4 spectroscopy observations \citep{Berne2025, Xuan2025}. However, if this candidate were confirmed as a true companion, along with others such as the candidate detected orbiting HD\,206893\,B \citep{2026A&A...705A.217K}, this would imply that binary gas giant planets may be more common than previously expected.

Follow-up imaging of Eps Ind A b using MIRI direct imaging would be advantageous over coronagraphy to determine whether this signal is caused by a systematic because of increased PSF stability, increased uniformity across the detector, and greater background sensitivity to cold companions.
Importantly, the primary advantage of MIRI imaging in this context is not necessarily improved sensitivity to lower-mass or tighter-separation companions, but rather the ability to robustly characterize instrumental systematics through well-calibrated PSF behavior. Future MIRI imaging observations could use archival data to empirically estimate a false-positive rate, since there is substantially more archival MIRI imaging data in the relevant bands than coronagraphic data. The extensive archival MIRI imaging dataset enables empirical assessment of PSF stability and residual structure, which is currently not possible for coronagraphic observations. This improved understanding is essential for distinguishing true astrophysical signals from systematic artifacts.

Future observations could also collect multiple observations of a reference star at different locations on the MIRI detector in order to build an even more robust effective point spread function (ePSF) \citep{AndersonKing2000, Anderson2016}, eliminating spatial PSF variation and interpolation artifacts as major sources of error.
Our lack of an ePSF means that the noise is correlated in both our target and reference images, leading to potentially increased correlations in the residuals. Furthermore, Eps~Ind~A~b was imaged very close to a 4QPM phase boundary, which makes modeling these correlations difficult due to interference from this boundary.

\subsection{Prospects for future exomoon/binary planet searches with direct imaging}

Direct imaging of exomoons is often dismissed as a viable detection method because the angular resolution required to probe orbital separations where moons are expected to be abundant implies the need for extremely large telescope apertures. For example, Callisto, the outermost Galilean moon, orbits at a separation of 0.013\,AU. Even at UV/visible wavelengths (e.g., $\lambda = 0.4\,\mu$m), resolving such a separation at a distance of a few parsecs would require a $\sim$20\,m telescope to achieve formal resolution (1$\lambda$/D), perhaps achievable with the upcoming generation of ELTs \citep{2021Msngr.182...38K}, but only with exquisite AO correction. 

However, as we have demonstrated here, it is possible to detect moons within the diffraction limit thanks to the extreme stability of space-based observatories and the use of double-PSF detection techniques.
This work lays the foundations for future exomoon searches using JWST/MIRI images of directly imaged companions in nearby systems, such as the exoplanet discoveries that may come from the GO 6122 \citep{CoolKidsprop}, GO 9056 \citep{Fransonjwstprop}, and SURVEY 8185 \citep{HOTHprop} programs. 
This result also suggests that such separations could be probed with smaller telescope apertures, such as the proposed Habitable Worlds Observatory \citep[HWO;][]{2021pdaa.book.....N}. Moreover, previous studies have shown that the contrast ratios between giant planets and terrestrial-sized moons can be comparable to those achieved in this work \citep{2015ApJ...812....5A}. Together, these results indicate that direct imaging of exomoons may represent a viable path forward for their detection despite their small separations.

\section{Conclusions} \label{sec:conc}

In this work, we demonstrate the capability of JWST to directly image binary planets/exomoons of nearby directly imaged exoplanets.
\begin{itemize}
    \item Using an injection–recovery test, we determine that our observations were sensitive to companions as small as 1.3\,M$_{\rm J}$ (T = 130\,K) at the outer edge of the Hill sphere, and 2.5\,M$_{\rm J}$ (T = 180\,K) at 0.52\,AU (0.3$\lambda/D$).
    \item Based on our contrast curve limits, we confidently rule out the presence of a similar-mass ($>$ 2.2 $M_{\mathrm{J}}$), large-separation ($>$ 0.8 AU) companion to Eps~Ind~A~b, barring a line-of-sight alignment at the time of observation. 
    \item Based on the detection limits achieved in this study, we conclude that mid-IR direct imaging with JWST/MIRI may represent a viable path forward for detecting and confirming binary planets and exomoons around the nearest directly imaged exoplanets.
    \item We detect a weak signal at an approximate separation of 0.16\arcsec\ and position angle of 285$^\circ$. We conclude that the signal is most likely due to systematics and not a companion. Future mid-infrared spectroscopy and/or imaging of this planet would be able to confirm or rule out if the residual pattern noted is instrumental or astrophysical in nature.
\end{itemize}

\begin{acknowledgments}
We thank the anonymous referee for their constructive comments, which improved the quality of this manuscript. This research made use of data hosted by the Mikulski Archive for Space Telescopes (\url{https://mast.stsci.edu/}).
MCG received support from the MIT Undergraduate Research Opportunities Program.
MDF is supported by an NSF Astronomy and Astrophysics Postdoctoral Fellowship under award AST-2303911. 
\end{acknowledgments}

\begin{contribution}

MCG was responsible for executing this project and writing/submitting this manuscript.
MAL and RBR ideated the initial project concept, served in advisory roles, and assisted MCG in preparing the manuscript. RBR also provided code to convert giant planet temperatures to masses.
MDF, an expert in double-PSF fitting, provided helpful advice and comments towards this end.
ECM, the original imager of Eps Ind A b, brought expertise on the system and helpful advice.
KF provided helpful advice and comments.
SCM served in an advisory role and helped procure funding.
LAP and AV also provided helpful advice and comments.

\end{contribution}

\facilities{JWST (MIRI)}

\software{
    astropy \citep{astropy1, astropy2, astropy3}, 
    corner \citep{corner},
    Matplotlib \citep{matplotlib},
    MultiNest \citep{multinest}
    NumPy \citep{numpy},
    PyMultiNest \citep{pymultinest},    
    SciPy \citep{scipy},
    StPSF \citep{Perrin_STPSF, PSF2012, PSF2014}
    VIP \citep{GomezGonzales2017, Christiaens2023}
}

\bibliography{manuscript}{}
\bibliographystyle{aasjournalv7}

\end{document}